\def\year{2019}\relax
\documentclass[letterpaper]{article} 
\usepackage{aaai19}  
\usepackage{times}  
\usepackage{helvet}  
\usepackage{courier}  
\usepackage{url}  
\usepackage{graphicx}  
\usepackage{amsmath}
\usepackage[linesnumbered, lined, boxed, ruled, vlined]{algorithm2e}
\usepackage{multirow}
\begin{document}
%
\title{A Novel Bin Design Problem and High Performance Algorithm for E-commerce Logistics System}
\author{Xinhang Zhang, Haoyuan Hu, Longfei Wang, Zhijun Sun, Ying Zhang, Kunpeng Han, Yinghui Xu \\
Zhejiang Cainiao Supply Chain Management Co., Ltd. 
}
\maketitle
\begin{abstract}
Packing cost accounts for a large part of the e-commerce logistics cost. 
Mining the patterns of customer orders and designing suitable packing bins help to reduce operating cost. 
In the classical bin packing problem, a given set of cuboid-shaped items should be packed into bins with given and fixed-sizes (length, width and height) to minimize the number of bins that are used. 
However, a novel bin design problem is proposed in this paper. 
The decision variables are the geometric sizes of bins, and the objective is to minimize the total surface area.  
To solve the problem, a low computational-complexity, high-performance heuristic algorithm based on dynamic programming and depth-first tree search, named DPTS, is developed. 
Based on real historical data that are collected from logistics scenario, numerical experiments show that the DPTS out-performed 5.8\% than the greedy local search (GLS) algorithm in the total cost. 
What's more, DPTS algorithm requires only about 1/50 times of the computational resources compared to the GLS algorithm. 
This demonstrates that DPTS algorithm is very efficient in bin design problem and can help logistics companies to make appropriate design.
\end{abstract}

\section{Introduction}

For logistics or production companies, packing cost accounts for a large portion of the total operating cost. 
So many optimization problems and algorithms have been studied and applied to reduce packing cost. 
One of the most classical and popular problems is three-dimensional bin packing problem. 
In this problem, a number of cuboid-shaped items with different sizes should be packed into bins orthogonally. 
The sizes and costs of bins are fixed and the objective is to minimize the number of bins used, i.e., minimize the total cost. 
This problem is NP-hard \cite{coffman1980performance} and has many applications in real world. 
So it has attracted many researchers' interest and some valuable achievements have been obtained.

In the classical bin packing problems, the sizes of bins are fixed and given. 
The decision variables to be optimized include packing sequence, items' orientations and positions. 
Effective and efficient bin packing algorithms can make good packing plans quickly and reduce the packing cost. 
But there is another important factor that should be considered: the sizes of bins. 
For example, if the bin is too large for most customer orders, packing space and materials will be wasted. 
On the other hand, if the bin is too small, then items from one customer order may be separated into several bins, which may lead to additional management and packing cost. 
For many e-commerce companies, a warehouse may manage and supply different categories of commodities, such as clothes, cosmetics and household electrical appliances. 
Commodities stored in each warehouse varies greatly in both size and category, so it is highly demanding that different warehouses should use different sized bins. 
Mining the patterns of size of customer orders using historical order data and designing different sized  bins for different warehouse operational scenario may save a lot of packing cost, which is the main motivation of study in this paper.

In this paper, a novel bin design problem is proposed. In this problem, packing bins with certain geometric sizes are designed based on the customer orders. 
Each customer order consists of several cuboid-shaped items. 
Items from one order must be put into the same bin while items from different orders cannot be put into one same bin. 
The number of bin types are given beforehand, and the decision variables to be optimized are the sizes (length, width and height) of bins. 
The cost of each bin is proportional to its surface area and the objective is to minimize the total cost of bins that are used to pack all customer orders.

\section{Related Work}
The most relevant research to the topic discussed in this paper is the studies on bin packing problems. 
Bin packing problems are first proposed in 1970s. 
Since then, related work about their variants, algorithms and applications has been researched. 
\cite{cote2014exact} develops a branch-and-cut algorithm to solve a two-dimensional orthogonal packing problem with unloading constraints. 
Lower bounds of the three-dimensional bin packing problem are discussed and an exact branch-and-bound algorithm, which can solve instances with up to 90 items, is proposed \cite{martello2000three}. 
In \cite{baldi2014branch}, a branch-and-price exact method and a beam search heuristics is proposed to solve the variable cost and size bin packing problem with optional items. 
Due to the difficulty of obtaining optimal solutions of bin packing problems, some approximation algorithms are investigated. 
\cite{baker198154} proposes a 5/4 algorithm for the two-dimensional packing problem. 
In \cite{scheithauer1991three}, an approximation algorithm for the three-dimensional bin packing problems is proposed and the performance is investigated. 
In \cite{seiden2003new}, the variable-sized online bin packing problem is studied and a new algorithm is proposed and its upper bound is analyzed. 
The first approximation scheme for a nontrivial two-dimensional generalization of a classical one-dimensional packing problem in which rectangles have to be packed in finite squares is studied in \cite{caprara2005fast}. 
And some effective heuristic algorithms, such as Tabu Search based algorithms \cite{lodi2002heuristic} \cite{lodi2004tspack} \cite{crainic2009ts}, genetic algorithms \cite{falkenauer1996hybrid} \cite{falkenauer1992genetic} \cite{reeves1996hybrid}, guided local search algorithm \cite{faroe2003guided}, ant colony based algorithm \cite{levine2004ant}, extreme point-based heuristics \cite{crainic2008extreme}, have been developed.

There is another branch of the packing problem: the strip packing problem. 
In this problem, some cuboid-shaped items should be packed into a given strip orthogonally. 
The length and width of the strip is fixed, and the objective is to minimize the height of packing. 
Some exact algorithms \cite{martello2003exact} \cite{kenmochi2009exact}, approximation algorithms \cite{steinberg1997strip} and heuristic algorithms \cite{bortfeldt2007heuristic} \cite{bortfeldt2006genetic} \cite{hopper2001review} have been investigated.

To the best of our knowledge, there is no research about optimizing the sizes of bins. 
The problem  introduced in this paper is novel and worth being studied.

\section{Bin Design Problem}
In this section, the definition of the novel bin design problem will be introduced. 
In the problem, a number of customer orders are given, each consists of several cuboid-shaped items.  
All the items belongs to each order should be packed into one specific bin. 
To guarantee customer experience, items from one order should be put into the same bin, and one packing bin only contains the items from the same order. 
The total number of bin types is limited and set beforehand due to the manufacturing cost, and the decision variables are the sizes (length, width and height) of bins. 
What's more, the sizes of bins in all three dimensions are monotonically increasing along the bin type index, therefore, the $i$th type of bin can be hold in the ($i$+1)th type of bin. 
The cost of each type of bin is proportional to its surface area. 
And the objective is to minimize the total cost to pack all customer orders.

To make our definition more clear and precise, some notations are introduced in Table 1.
\begin{table}
\newcommand{\tabincell}[2]{\begin{tabular}{@{}#1@{}}#2\end{tabular}}
  \caption{Notations in Bin Design Problem}
  \label{tab:notations}
  \begin{tabular}{c|l}
  \hline
    Notation&Meaning\\  \hline
    $N$ & The number of customer orders \\ 
    $K$ & The number of types of bins  \\
    $l_k$ & The length of the $k$th type of bin \\
    $w_k$ & The width of the $k$th type of bin \\
    $h_k$ & The height of the $k$th type of bin \\
    $L$ & The upper bound of length of bins \\
    $W$ & The upper bound of width of bins \\
    $H$ &The upper bound of height of bins \\
    $F(l_k, w_k, h_k)$ & \tabincell{l} {   
    Number of customer orders that can be \\ packed  into the $k$th type of bin.}\\ \hline
\end{tabular}
\end{table}

Based on the description and notations above, the formulation of the bin design problem is presented as follows:

\begin{small}
\begin{align}
& {\text{min}}
& & \sum_{i = 1}^{K + 1} c_k (F(l_k, w_k, h_k) -  F(l_{k-1}, w_{k-1}, h_{k-1})) \\
& \text{s.t.} & & l_k \geq l_{k-1}, \forall k \in \{2, \dots, K \} \\
& & &  w_k \geq w_{k-1}, \forall k \in \{ 2, \dots, K \} \\
& & &  h_k \geq h_{k-1}, \forall k \in \{2, \dots, K\} \\
& & &  l_k + w_k + h_k > l_{k-1} +  w_{k - 1} + h_{k - 1}, \forall k \in \{2, \dots, K \} \\
& & &  l_k \leq L, \forall k \in \{1, 2, \dots, K\} \\
& & &  w_k \leq W, \forall k \in \{1, 2, \dots, K\} \\
& & &  h_k \leq H, \forall k \in \{1, 2, \dots, K\} \\
& & &  l_k, w_k, h_k \in Z^{+},\forall k \in \{1, 2, \dots, K\}  
\end{align}
\end{small}

The objective function indicates that the optimization goal is to minimize the total cost of bins that are used to pack all orders. For $k \leq K$, the cost of each of the $k$th type of bin with length $l_k$, width $w_k$ and height $h_k$ is $c_k = 2(l_k w_k+w_k h_k+ l_k h_k)$, and for $k = K + 1$, $c_k = +\infty, F(l_k, w_k, h_k) = N$. The number of the $k$th type of bins that are used is $F(l_k, w_k, h_k) - F(l_{k-1}, w_{k-1}, h_{k-1})$, because if a customer order can be packed into the $(k - 1)$th type and the $k$th type of bin, it must be packed into the  $(k -1)$th type of bin, because the surface area of the  $(k - 1)$th type of bin is smaller than that of the $k$th type of bin.

Constraints (2)-(5) denote that the $(k - 1)$th type of bin can be contained in the  $k$th type of bin. Constraints (6)-(9) indicate that the sizes of the bins are positive integers and not greater than the corresponding upper bounds.

To solve the problem above, $F(l_k, w_k, h_k)$ for each combination of $l_k$, $w_k$ and $h_k$ must be calculated first. What's more, the objective function is nonlinear, so traditional algorithms are not capable of solving the problem directly. In next section, a heuristic algorithm based on dynamic programming and depth-first tree search is developed to solve the problem. 

\section{Heuristic Algorithm}
In this section, a heuristic algorithm based on dynamic programming and depth-first tree search, named DPTS algorithm,  will be discussed. The dynamic programming method is used to find the optimal combinations of bin types if $F(l_k, w_k, h_k)$ for each type of bins has been obtained. Depth-first tree search method is used to search the marginal bin type for each customer order and $F(l_k, w_k, h_k)$ can be calculated based on the search results. The framework of DPTS is showed in Fig. 1.

\begin{figure}
\begin{center}
\includegraphics[height=2.8in, width=1.0in]{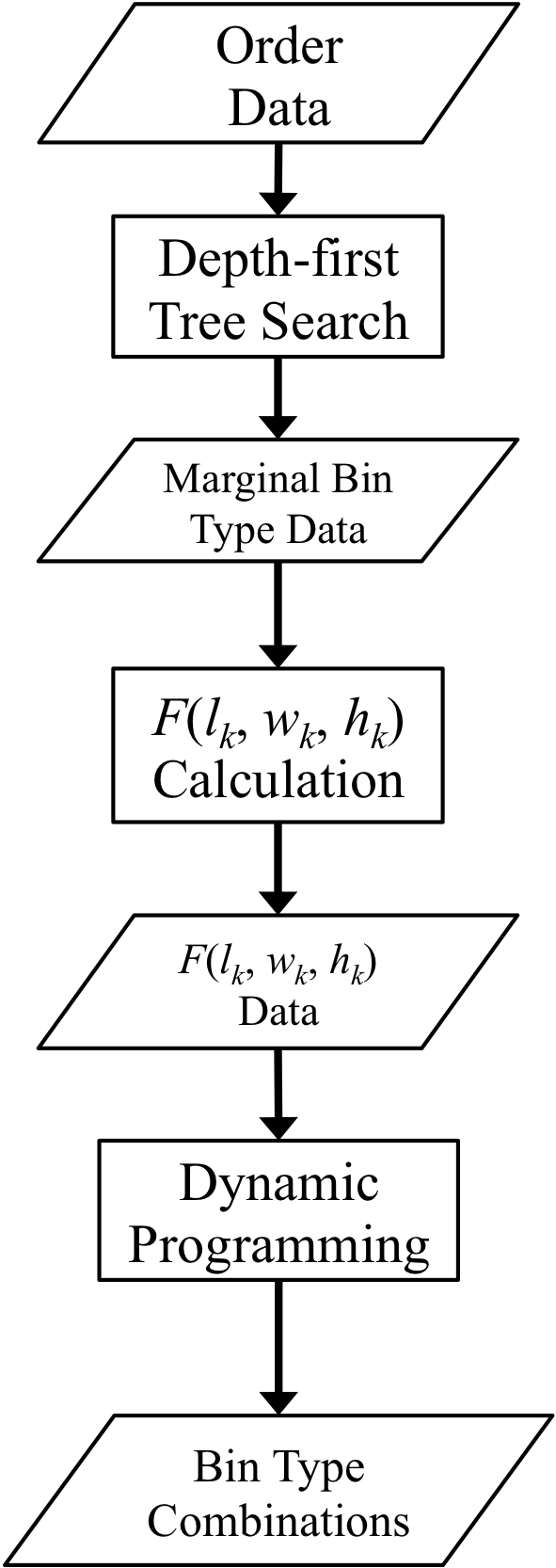}
\caption{DPTS Framework}
\end{center}
\end{figure}

For the convenience of the discussion, some definitions will be introduced first.

Definition 1.  A bin type with $(l_k, w_k, h_k)$ is an expanded bin type for a bin type with  $(l_l, w_l, h_l)$ if $l_k \geq l_l, w_k \geq w_l, h_k \geq h_l$, $(l_k + w_k + h_k) > (l_l+w_l+h_l),$ and bin type with $(l_l, w_l, h_l)$ is a shrunken bin type for bin type with  $(l_k, w_k, h_k)$.

Definition 2. A bin type with $(l_k, w_k, h_k)$ is a marginal bin type for a customer order if the order can be packed into this type of bin, but cannot be packed into any other bin that is shrunken type of this bin type.

\subsection{Dynamic Programming Method}

Given $F(l_k, w_k, h_k)$ for each combinations of $l_k$, $w_k$ and $h_k$, the problem proposed in the above section can be solved by a dynamic programming method. As shown in Fig. 2, there are  $(K+2)$ decision steps. For the $k$th step, where $k \in \{1, 2, \dots, K\}$, the sizes of the $k$th type of bins are selected from $L\times W \times H$ candidates. Let $b_k$ denote the type of bin that is selected in the $k$th step, $b_k=(l_k, w_k, h_k)$, and let $C(k, b_k)$ denote the total cost if $b_k$ is selected in the $k$th step. $I(b_{k-1}, b_k)$ is used to indicate the incremental cost if $b_{k-1}$ is selected in the $(k-1)$th step and $b_k$ is selected in the $k$th step. Then the recursive function is:

$C(k, b_k) = \min_{b_{k - 1}}  \{ C(k-1, b_{k-1})+ I(b_{k-1}, b_k) \} $

where 
\begin{small}
$$
I(b_{k - 1}, b_k) = 
\left \{
  \begin{aligned}
  cost(b_k)(F(b_k) - F(b_{k - 1})) & , if ~ b_k ~ is ~ an ~ expanded \\
  &  ~ bin ~ type ~ of ~  b_{k-1} \\
  +\infty & ,otherwise
  \end{aligned}
  \right.
$$

$cost(b_k)=2(l_k w_k +w_k h_k + l _k h_k ),\forall k  \in \{1, 2, \dots, K\}$

$b_0=(0,0,0), F(b_0)=0, C(0,b_0)=0$

$b_{K+1} =(L+1,W+1,H+1), F(b_{K+1})=N, cost(b_{K+1})= + \infty$
\end{small}

The dynamic programming method above can be interpreted as finding the shortest path between the point in step 0 and the point in step $(K+1)$. The definition of  $I(b_{k-1}, b_k)$ can assure that $b_k$ must be an expanded type for $b_{k-1}$ in the final solution. And because $cost(b_{K+1})= + \infty$, so $F(b_K)$ must equal to $N$ in the final solution, i.e., the $K$th type of bin must be able to pack all customer orders.

The computation complexity of the dynamic programming method is $O(K \times L^2 \times W^2 \times H^2)$. To accelerate the computation, some tricky techniques based on ideas of divide and conquer  and convex hull are used, and the computation complexity is reduced to $O(K \times L \times W \times H \times log(L) \times log(W) \times log(H))$. The accelerative dynamic programming algorithm can be seen in the Appendix.

\begin{figure}
\begin{center}
\includegraphics[height=1.6in, width=2.8in]{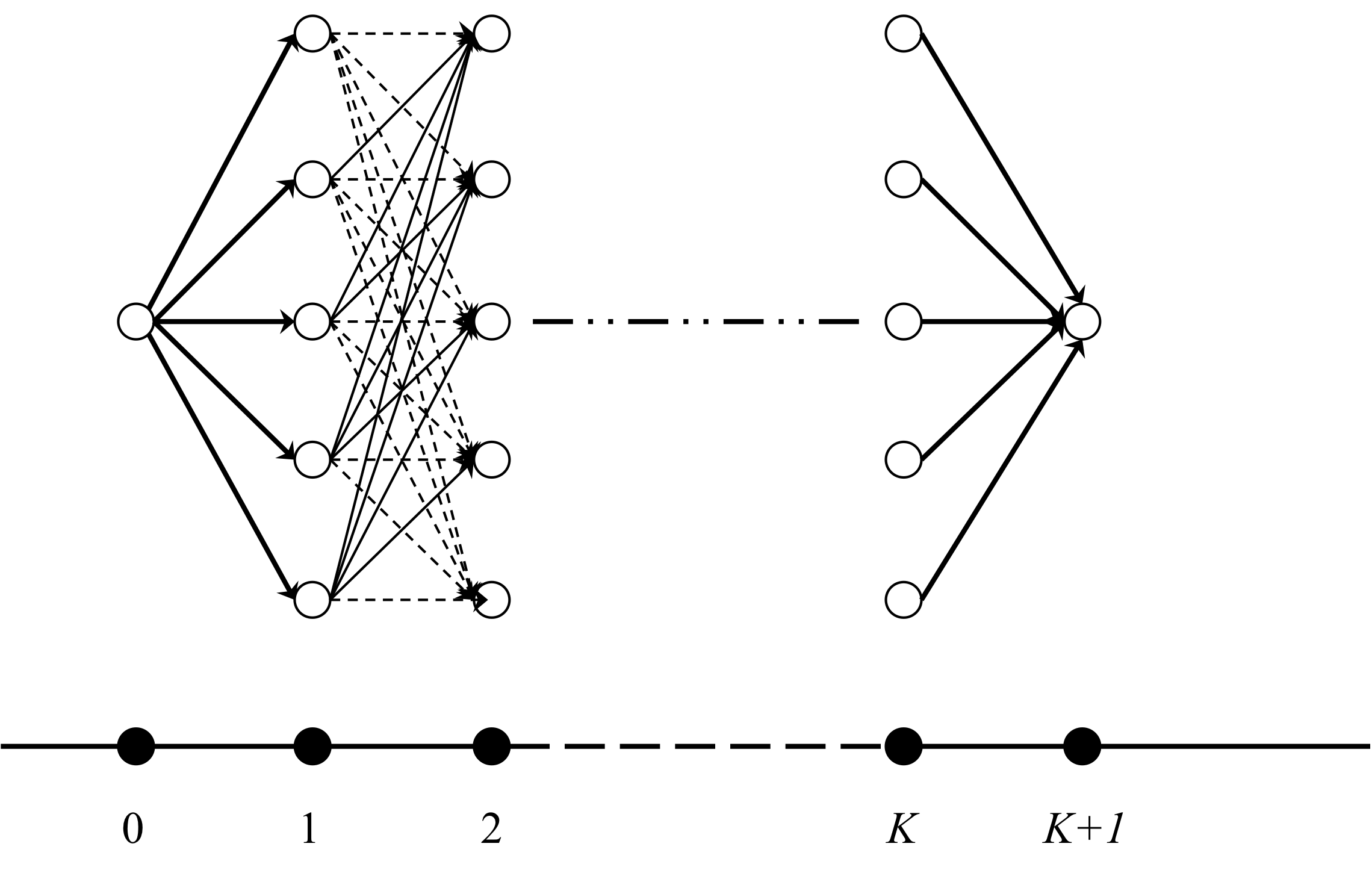}
\caption{Dynamic Programming Method}
\end{center}
\end{figure}

\subsection{Tree Search Method}

To calculate $F(b_k)$ for each bin type $b_k$, one simple idea is counting the number of customer orders that can be put into the bin type, this needs an effective three dimensional bin packing algorithm to determine whether a customer order can be put into a bin or not. When the number of customer orders and the number of candidate bin types is large, the computation cost will be huge. So in this section, a depth-first tree search method is designed and implemented to calculate $F(b_k)$.

The depth-first tree search method is to search the marginal bin types for each customer order, which will be used to calculate $F(b_k)$. The basic idea of the tree search method is to sort the items in one customer order in a descending order of volume first, then search the orientation and position of each item. For the convenience of the discussion of the tree search method, some notations are introduced below.

ItemOrientation: the orientation set for each item, which is a subset of {front-up, front-down, side-up, side-down, bottom-up, bottom-down}. Because some items have same size at different dimensions, the size of orientation set may be less than 6, which can reduce the computation cost.

CornersX, CornersY, CornersZ: the dictionary of x, y, z-coordinates of corner points of the items and the left-bottom-back corner point of bin. The keys of the dictionary are the coordinates and the values are the number of times that each coordinate appears.

When putting a new item, it is assumed that the position of the left-bottom-back corner point of the item must be a combination of x-coordinate, y-coordinate, z-coordinate from CornersX, CornersY, CornersZ, respectively. A two-dimensional example is show in Fig. 3. Before putting items, the initial CornersX = $\{0:1\}$, CornersY = $\{0:1\}$. And after putting item 1, CornersX is updated to $\{0:1, 2:1 \}$, CornersY is updated to $\{0:1, 1:1\}$, after putting item 2, CornersX=$\{0:1, 1:1, 2:1\}$, CornersY=$\{0:1, 1:1, 2:1\}$.

\begin{figure}
\begin{center}
\includegraphics[height=1.6in, width=1.6in]{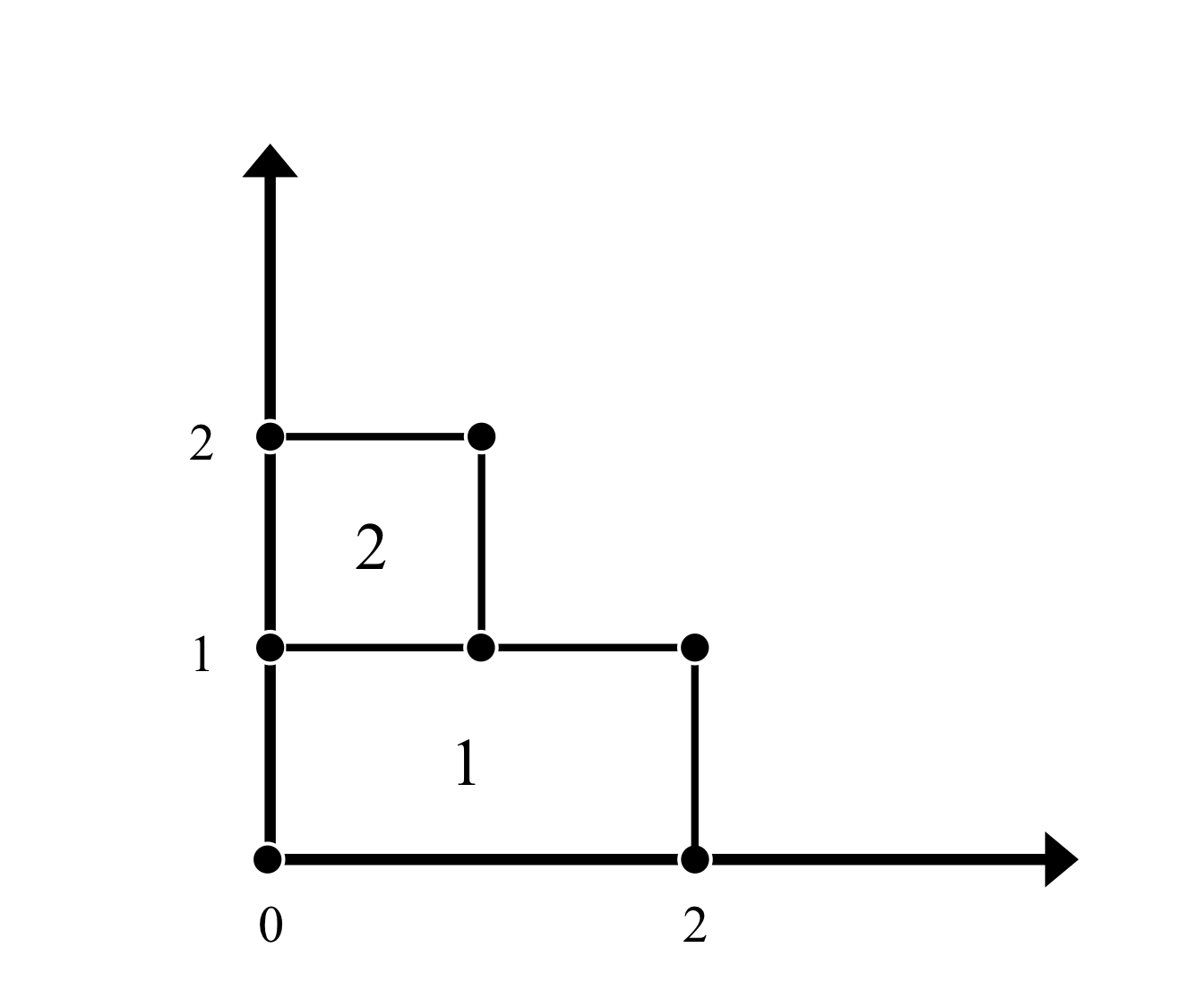}
\caption{Example of Corners Dictionary}
\end{center}
\end{figure}

BestCost: the lowest cost of leaf node that is found in the search process.

NodeBinType: the length, width and height of the smallest bin that can contain items in the node.

Based on the notations above, the depth-first tree search method is shown in Algorithm 1 and Algorithm 2. In Step 8 of the DepthFirstTreeSerach, the algorithm enumerates every possible orientation and position to place a new item. In Step 11, the algorithm should determine whether a boundary and an item overlap, and whether the item is next to the boundary on x-axis, y-axis or z-axis. Being next to the boundary on x-axis means that the item cannot be moved in the direction to decrease x-coordinate of corner points of the item. In Step 14, the algorithm prunes some branches, because the marginal bin type won't have a large cost.

\begin{algorithm}
\DontPrintSemicolon
 \Begin{
 $BestCost$ $\longleftarrow + \infty$\;
 $MarginalBinTypeSet$ $\longleftarrow \emptyset$\;
 $totalSearchCount$ $\longleftarrow 0$\;
 $CornersX$   $\longleftarrow \{0:1\}$\;
 $CornersY$   $\longleftarrow \{0:1\}$\;
 $CornersZ$   $\longleftarrow \{0:1\}$\;
 $RootNodeBinType$ = $(0, 0, 0)$\;
 $M$ $\longleftarrow$ the number of items
 
 Sort items in a descending order of volume, and get sorted index of each item\;
 Calculate ItemOrientation for each item\;
 \textbf{DepthFirstTreeSearch}(0, RootNodeBinType, 0)\;
 
 \For{$(l, w, h) \in$ MarginalBinTypeSet} {
 	\If{$(l, w, h)$  is a shrunken bin type for any other bin type in MarginalBinTypeSet}{	
		Remove $(l, w, h)$ from $MarginalBinTypeSet$ \;
		}
	}
 }
\caption{MarginalBinTypeSearch\label{IR}}
\end{algorithm}

\begin{algorithm}
\DontPrintSemicolon
\KwData{itemIndex, NodeBinType, NodeCost}
 \Begin{
 \If{itemIndex = M}{
 	$BestCost = \min(BestCost, NodeCost)$\;
	$MarginalBinTypeSet$.add(NodeBinType)\;	
 }
 $totalSearchCount = totalSearchCount + 1$\;
 \If{totalSearchCount $>$ maxSearchCount} {
 	return\;
 }
 
  \For{orie $\in$ ItemOrientation(itemIndex), $x \in $ CornersX.keySet, $y \in $ CornersY.keySet, $z \in$ CornersZ.keySet} {
 	Let $(x, y, z)$ be the coordinates of the left-bottom-back corner point of the item \;
	Calculate the coordinates of the right-upper-front corner point, which is denoted as $(x', y', z')$\;
	\If{not\_out\_bound and not\_overlap and next\_to\_boundary\_on\_every\_axis} {
		newBinType = ($\max$($x'$, NodeBinType($x$)), $\max$($y'$, NodeBinType($y$)), $\max$($z'$, NodeBinType($z$)))\;
		newCost = 2 (newBinType($x$) $\times$ newBinType($y$) + newBinType($x$) $\times$ newBinType($z$) + newBinType($y$) $\times$ newBinType($z$)) \;
		\If{newCost $\leq$ BestCost $\times$ $Z$} {
			addToCorner($x'$, $y'$, $z'$)\;
			\textbf{DepthFirstTreeSearch}(itemIndex + 1, newBinType, newCost)\;
			removeFromCorner($x'$, $y'$, $z'$)\;
		}	
	}
 }
}
 \caption{DepthFirstTreeSearch\label{IR}}
\end{algorithm}

The running time of the tree search algorithm depends on the number of items in customer order. And customer orders are independent, so we can execute the tree search in parallel to reduce total computation time.

To calculate $F(l, w, h)$, we define the intermediate results $f(l, w, h)$, where $F(l, w, h)=\sum_{i=0}^l \sum_{j=0}^w f(i, j, h)$ for each $l \in \{1, 2, \dots, L\}, w \in \{1, 2, \dots, W\},h \in \{1,2, \dots,H \}$. The algorithm to calculate $f(l, w, h)$ is shown in Algorithm 3.

In Algorithm 3, the computation of $f(i, j, h)$ for each $h \in \{0,1,2, \dots, H\}$ is also independent, so the computation can be executed in parallel to reduce total computation time.

Lemma 1. Obtaining $f(l, w, h)$ from the Algorithm 3, then $F(l, w, h)$ can be calculated by $F(l, w, h)=\sum_{i  = 0}^l \sum_{j = 0}^w f(i, j, h)$.

Proof: The lemma can be proved by a mathematical induction method.

Firstly, the statement in the Lemma is obviously holds for $N=0$, i.e., when the number of customer orders is zero.

Secondly, assume that the statement holds when the number of orders is $N$, then a new order is added, and assume the marginal bin types of the $(N+1)$th order is $M_{N+1}  =\{(l_{i1}, w_{i1}, h_{i1}),\dots,$
$(l_{im_{N+1}}, w_{im_{N+1}}, h_{im_{N+1}} )\}$.

Based on the logic of Algorithm 3, for any $F(l, w, h)$, it will be affected by the new order only if there is bin type in $M_{N+1}$ that is shrunken bin type of bin type $(l, w, h)$. The shrunken bin types are processed as the Step 11 $-$ Step 15 in Algorithm 3, then we can get the a set of $\{(l_{i1}, w_{i1}, h), (l_{i2}, w_{i2}, h), \dots, (l_{im'_{N+1}}, w_{im'_{N+1}}, h) \}$.

And $F(l, w, h)$ will be affected by the set, because $f(i, j, h)$ will be changed by the set. Let $F'(l, w, h)$ denote the new $F$ for $(l, w, h)$, and let $f'(i, j, h)$ denote the new $f$, then:

\begin{small}
$F'(l, w, h) - F(l, w, h) =(f'(l_{i1}, w_{i1}, h) - f(l_{i1},w_{i1}, h) + \dots + $
$f'(l_{im'_{N+1}}, w_{im'_{N+1}}, h) - f(l_{im'_{N+1}}, w_{im'_{N+1}}, h)) + (f'(l_{i2}, w_{i1}, h) - f(l_{i2}, w_{i1}, h) + \cdots +
f'(l_{im'_{N+1}}, w_{i, m'_{N+1} - 1}, h) - f(l_{im'_{N+1}}, w_{i,m'_{N+1} - 1}, h))
=\sum_{i = 1}^{m'_{N+1}} 1 + \sum_{i = 1}^{m'_{N+1} - 1} (-1) = 1$
\end{small}

So after a new order is added, and if the new order can be packed into any bin type $(l, w, h)$ (there is at least one marginal bin type of the order that is shrunken bin type of bin type $(l, w, h)$), $F(l, w, h)$ will be increased by one.  This satisfies the definition of $F(l, w, h)$. This ends the proof.

\begin{algorithm}
\DontPrintSemicolon
 \Begin{
 Initialize $f(l, w, h) = 0, \forall l \in \{1, 2, \dots, L\}, w \in \{1,2, \dots, W \},h \in \{1, 2, \dots ,H\}$. Let $M_i$ be the marginal bin type set of the $i$th  customer order, let $m_i=|M_i|$, and let $(l_{ij}, w_{ij}, h_{ij})$ be the length, width and height of the $j$th marginal bin type in $M_i$\;
 
 \For{$h \in \{0, 1, \dots, H \}$} {
 	\For{$i \in \{1, 2, \dots, N \}$} {
		$M'_i = \emptyset$ \;
		\For{ $j \in \{1, 2, , \dots, m_i\}$} {
			\If{$h_{ij} > h$} {
				continue \;
			} \Else {
				$M'_i = M'_i \cup \{(l_{ij}, w_{ij}, h) \}$ \;
			}
		}
		
		\If{$M'_i \neq \emptyset$} {
			\For{$(l ,w, h) \in M'_i$} {
				\If{$(l, w, h)$ is a shrunken bin type of any other bin type in $M'_i$} {
					Remove $(l, w, h)$ from $M'_i$ \;
				}
			}
			
			Sort $(l, w, h)$ in $M'_i$ in an ascending order of $l$, assume the sorted $M'_i$ is $\{(l_{i1}, w_{i1}, h), (l_{i2}, w_{i2}, h ), \dots, (l_{im'_i}, w_{im'_i}, h)\}$
			
			\For{$j \in \{1, 2, \dots, m'_i \}$} {
				\If{$j = 1$} {
					$f(l_{ij},w_{ij},h)  = f(l_{ij},w_{ij},h)  + 1$ \;
				} \Else {
					$f(l_{ij}, w_{ij},h)   = f(l_{ij},w_{ij},h)  + 1$ \;
					$f(l_{ij}, w_{i, j - 1},h) = f(l_{ij},w_{i, j - 1},h)  - 1$\;
				}
			}
		}
	}
 }
 }
  \caption{Calculate $f(i, j, h)$ \label{IR}}
\end{algorithm}

Based on the lemma above, after obtaining $f(l, w, h)$, we can calculate $F(l, w, h)= \sum_{i = 0}^l \sum_{j = 0}^w f(i, j, h)$. To reduce computation cost, we can get the following recursive function based on the Inclusion-exclusion principle:

$F(l, w, h)=F(l-1, w, h) + F(l, w-1, h) - F(l-1, w-1, h) + f(l, w, h)$.

\section{Numerical Experiments}

\begin{figure}
 \begin{center}
	\includegraphics[scale=0.4]{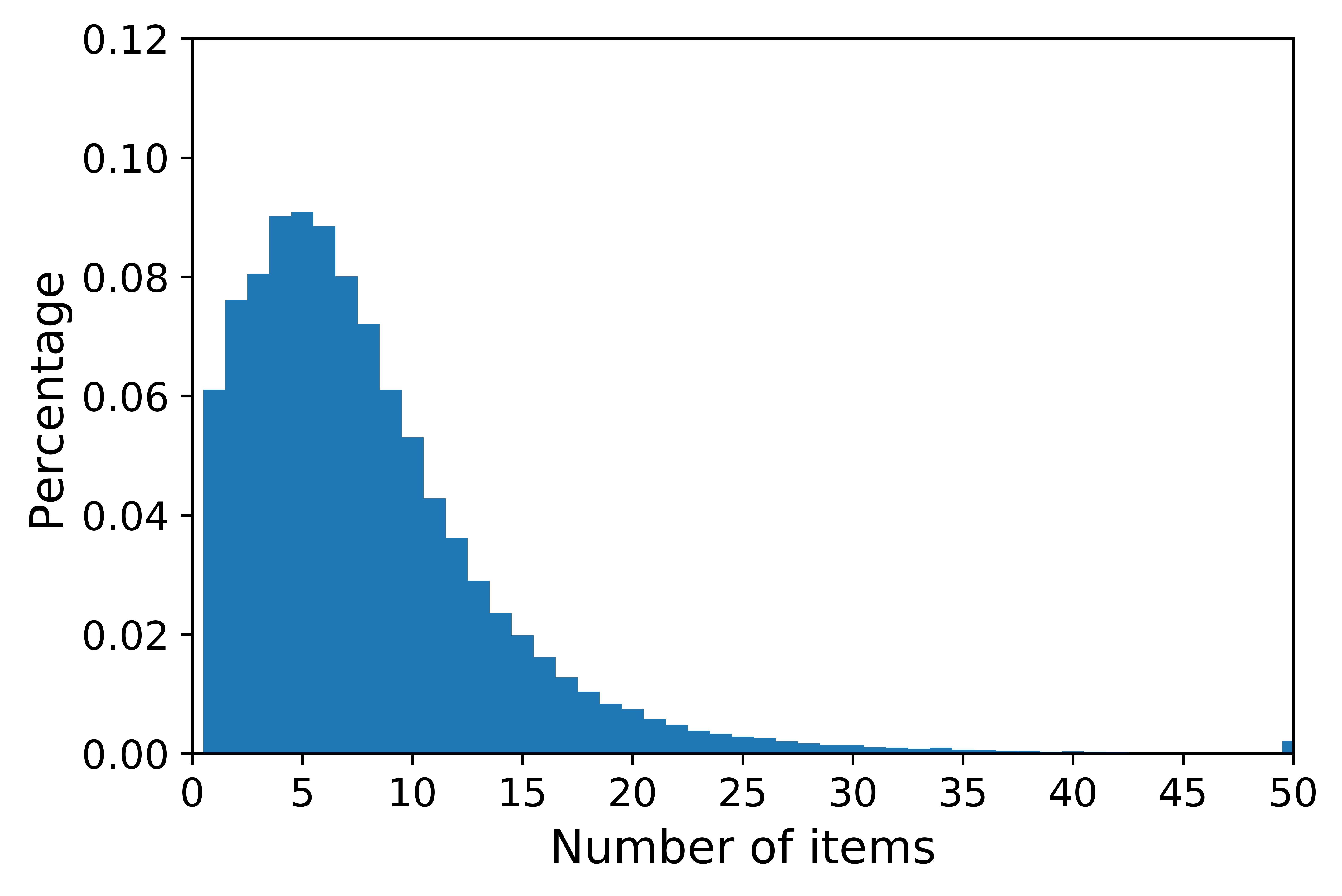}
	\caption{Probability distribution of item numbers within an order}
	\label{fig:item_number}
\end{center}
\end{figure}

To demonstrate performance of the DPTS algorithm proposed in this paper, numerical experiments are designed and conducted. 
The order data in experiments are all collected from real e-commerce business. 
The distribution of item number in orders is shown in  Fig. \ref{fig:item_number}. It can be seen that about 75 \% orders have less than or equal to 10 items. 
The experiments is classified into three categories based on the number of customer orders, i.e., 200000, 500000 and 1000000. 

A greedy local search (GLS) algorithm and decisions of human (purchasing expert who is responsible for the final plan) are used to compare to the heuristic algorithm. The basic idea of the GLS algorithm is to gradually enlarge or shrink the bin size at every step, and then invoke a three-dimensional bin packing algorithm to evaluate the combination of bin types. If the new combination of bin types can result in less packing cost, it will be accepted and its neighbors will be searched in next step.  The details of the GLS algorithm in Algorithm 4.

\begin{figure}[ht]
 \begin{center}
	\includegraphics[scale=0.4]{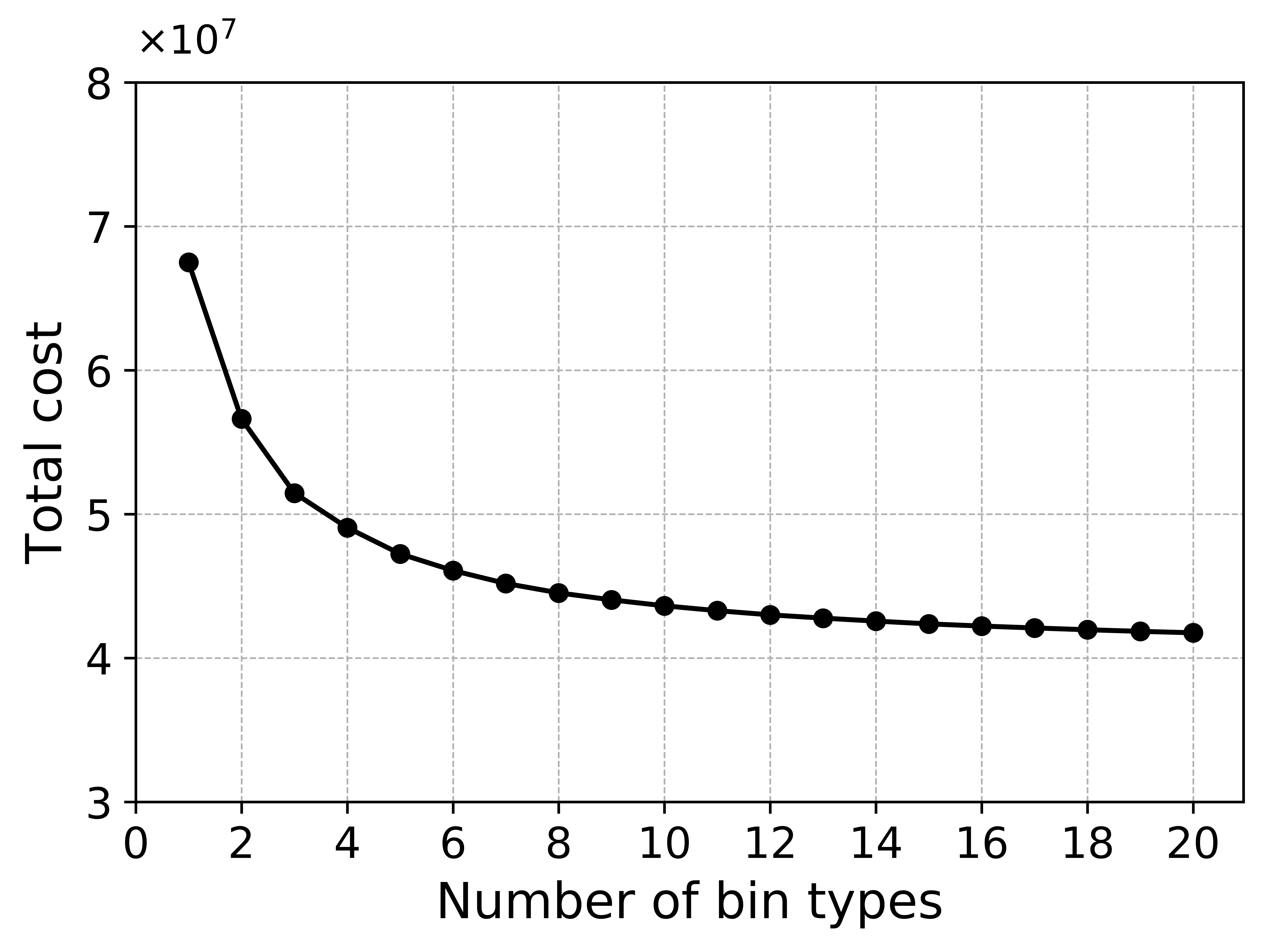}
	\caption{Total cost at different number of bin types}
	\label{fig:total_cost}
\end{center}
\end{figure}

\begin{figure}[ht]
  \begin{center}
	\includegraphics[scale=0.4]{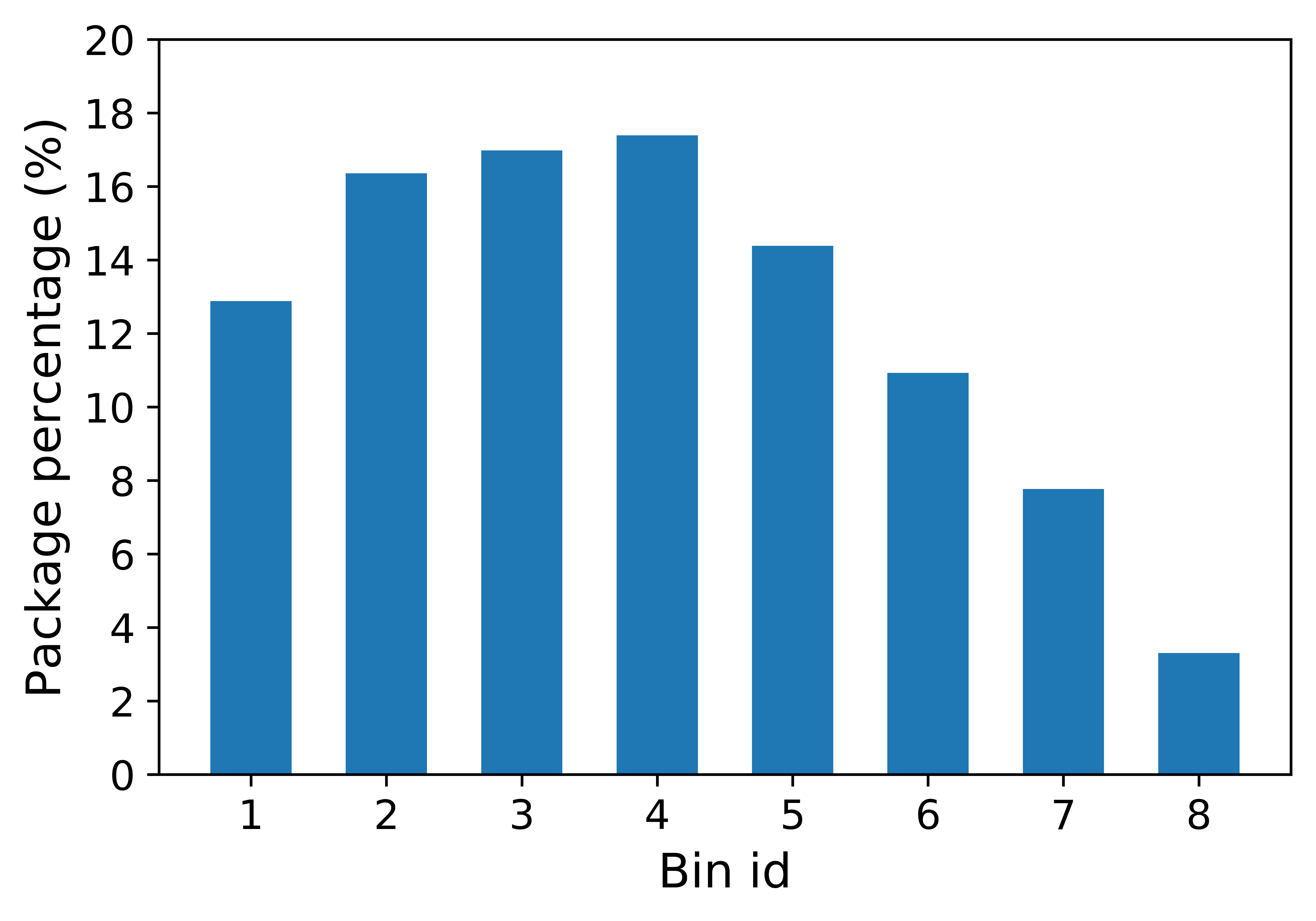}
	\caption{Order percentage of each bin type when $K = 8$}
	\label{fig:package_percentage}
\end{center}
\end{figure}

DPTS can optimize the total cost for different bin type numbers, and the results for 1000000 orders are shown in Fig. \ref{fig:total_cost}.
The number of bin types is usually set to be 8 in the usual practice of bin purchasing, so the detail sizes of 8 bin types are shown in Table 2 and the percentages of orders that can be packed into each bin type are shown in Fig. \ref{fig:package_percentage}.  The cost and computational time comparison are shown in Table~\ref{tab:results}. DPTS reduce cost by 12.8\% compare to human result, and 5.8\% compare to GLS. More importantly, the DPTS has been used in real business and financial analysis result show that the algorithm can save about \$ 20 million packing cost a year (about 10\% of the total packing cost). What's more, DPTS algorithm is about 50 times faster than the GLS algorithm. In practice, the decisions about bin purchasing for many warehouses should be made in finite time, so the efficiency of DPTS is a big advantage in applications.

\begin{algorithm}
\DontPrintSemicolon
 \Begin{
 Select the bin types that are designed by warehouse managers as initial solution, and sort the bins by an increasing order \;
 $C \longleftarrow 0$ \;
 \For{$i \in \{1, 2, \dots, N \}$} {
 	\For{$l \in \{1, 2, \dots, K\}$} {
		\If{$(l_k, w_k, h_k)$ can pack the $i$th order} {
			$C = C + 2(l_k w_k + w_k h_k + l_k h_k)$ \;
			break;
		}
	}
 }
  
 NonImprovementCounter $\longleftarrow$ 0 \;
 \While{NonImprovementCounter $\leq$ NonImprovementThreshold} {
 Randomly select a dimension $i$ \;
 Randomly select a bin type $k$ \;
 Let $x_j, x_{j - 1}, x_{j + 1}$ be the size of dimension $i$ of bin type of $j, j - 1, j + 1$ respectively \;
 Randomly select the search direction, i.e., increase or decrease $x_j$ by one step size, obtain $x'_j$ \;
 
 \If{$x'_j < x_{j - 1}$ or $x'_j > x_{j + 1}$} {
 	continue \;
 } \Else {
 	Calculate new packing cost $C'$ \;
	\If{$C' < C$} {
		Use $x'_j$ to update the $k$th bin type \;
	} \Else {
		NonImprovementCounter = NonImprovementCounter + 1 \;
	}
 }
 	
 }
 }
  \caption{Greedy Local Search}
\end{algorithm}

 \begin{table}
 \centering
 \small
  \caption{Size of Bin Types}
  \label{tab:size}
  \begin{tabular}{c|c c c c c c c c}
    \hline
    {Bin id} & 1 & 2 & 3 & 4 & 5 & 6 & 7 & 8 \\ \hline
    L(cm) & 27 & 31 & 35 & 40 & 40 & 43 & 44 & 50  \\
    W(cm) & 18 & 23 & 25 & 28 & 28 & 30 & 35 & 40 \\
    H(cm) & 15 & 18 & 20 & 20 & 25 & 27& 30 & 33 \\ \hline
\end{tabular}
\end{table}

 \begin{table}
 \footnotesize
\newcommand{\tabincell}[2]{\begin{tabular}{@{}#1@{}}#2\end{tabular}}
  \caption{Comparison of performance in Bin Design Problem}
  \label{tab:results}
  \begin{tabular}{|c|c|c|c|c|c|}
    \hline
    \multirow{2}{*}{\tabincell{c}{No. of \\Orders}} & \multicolumn{2}{c|}{DPTS} & \multicolumn{2}{c|}{GLS} & Human \\ \cline{2-6}
    & \tabincell{c}{Total Cost \\ ($m^2$)} & \tabincell{cc}{CPU \\ Time \\ (s)} & \tabincell{c}{Total Cost \\ ($m^2$)} & \tabincell{cc}{CPU \\ Time \\ (s)} & \tabincell{c}{Total Cost \\ ($m^2$)} \\ \hline
    $2 \times 10^5$ & $8.811 \times 10^4$ & 2282 & $9.365 \times 10^4$ & $> 10^5$ & $1.006 \times 10^5$ \\ \hline
    $5 \times 10^5$ & $2.202 \times 10^5$ & 3390 & $2.340 \times 10^5$ & $> 10^5$ & $2.514 \times 10^5$ \\ \hline
    $1 \times 10^6$ & $4.453 \times 10^5$ & 8176 & $4.729 \times 10^5$ & $> 10^5$ & $5.109 \times 10^5$ \\ \hline
\end{tabular}
\end{table}

\section{Future Work}

The tree search algorithm simply traverse every orientation and position regardless of the potential reward at each search branch. In the future work, we may use a function to evaluate the fitness of every orientation and position, and search in sorted fitness order to converge faster, thus more branches can be pruned to accelerate the algorithm. And for orders with more items, we may use distributed computing to reduce computation time.

\section{Conclusion}

In this paper, a novel bin design problem is proposed. Different from the classical bin packing problems, the decision variables to be optimized are the sizes of bins that are used to pack customer orders, and the objective is to minimize the total surface area of bins. Due to the complexity of the problem, a high-performance heuristic algorithm based on dynamic programming and depth-first tree search is developed. In the algorithm, depth-first tree search algorithm is used to search marginal bin types for each customer order, and based on search results, the number of customer orders that can be packed for each type of bin can be calculated, then a dynamic programming method can be applied to obtained the optimal combination of bin types. Numerical experiments' results show that the heuristic algorithm outperforms a greedy search algorithm in terms of quality and efficiency. And financial analysis result show that the algorithm can save about \$ 20 million a year. Our main contributions include: firstly, a novel bin design problem is proposed; secondly, an elaborate and effective heuristic algorithm is designed and implemented to solve the problem and numerical experiments and financial analysis based on real data are conducted to demonstrate the value of the algorithm.

\section{Appendix}

\subsection{The Idea of Divide and Conquer}

For a dynamic programming function as follows:
$DP(k, j) = \min_{i} \{DP(k - 1, i) + g(i, j)\}$,  where $i \leq j$.

We define $U(M,Q)$ as the time complexity of updating $Q$ values in the $k$th stage by $M$ values in the $(k-1)$th stage, if the time complexity of each update is $U(1, 1)$, then the total complexity is $Q \times M \times U(1, 1)$.

The basic process of divide and conquer method to update states is shown in Algorithm 5.

\begin{algorithm}
\DontPrintSemicolon
 \Begin{
 \If{left $<$ right} {
 	middle = (left + right) / 2 \;
	DivideConquer(left, middle) \;
	DivideConquer(middle + 1, right) \;
	Update $DP(k, j)$ where $j \in $ \{middle + 1, $\dots$, right\} by $DP(k - 1, i)$ where $i \in $ \{left, \dots, middle\} \;
 } else {
 	Update $DP(k, right)$ by $DP(k - 1, left)$ \;
 }
 }
  \caption{DivideConquer\label{IR}}
\end{algorithm}

The recursive process will have $logN$ levels, on level $k (1 \leq k \leq logN) $, we update $2^{k-1}$ times, each update has time complexity $U(N/2^k, N/2^k)$, because on level $k$, $M = Q = N/2^k$. If we can reduce $U(M, Q)$ from $Q \times M \times U(1, 1)$ to $(M + Q) \times U(1, 1)$, the total time complexity of the process will be $1 \times (N/2 + N/2) \times U(1,1) + 2 \times (N/4+N/4) \times U(1,1)+ \cdots + N/2 \times (1+1) \times U(1,1) = N \times logN \times U(1,1)$. The method to reduce time complexity will be discussed in next.

If we store the update information as a list of ${(a, b) \rightarrow (c, d)}$, which means that updating $DP(k, j)$ where $j \in \{c, \dots, d\}$ by  $DP(k-1, i)$ where $i \in \{a, \dots, b\}$, then the time complexity will be $\sum_{x \in \{1, \dots, logL\}} \sum_{y \in \{1, \dots, logW\}}$

$\sum_{z \in \{1, \dots, logH\}} (2^{x -1} \times 2^{y - 1} \times 2^{z - 1} \times U(\frac{L}{2^x} \times \frac{W}{2^y} \times \frac{H}{2^z}, \frac{L}{2^x} \times \frac{W}{2^y} \times \frac{H}{2^z})) = L \times W \times H \times log(L) \times log(W) \times log(H) \times U(1, 1)$.

\subsection{The Idea of Convex Hull}

Because the dynamic programming function in Section Dynamic Programming Method is a linear function with one independent variable ($cost(b_k)$), so the process of updating $Q$ values in the $k$th stage by $M$ values in the $(k-1)$th stage is equivalent to the following process: for each of the $Q$ values, find the minimum output value from $M$ linear functions. The simple algorithm is to calculate the output value of each function and identify the minimum result. The total time complexity is $O(M \times Q)$. In this part, we show that the time complexity can be reduced to $O(M+Q)$.

Let $y_i=a_i x+b_i$ indicate the $i$th linear function, where $i \in \{1,\dots, M\}$. As show in Fig. 7, we can calculate some intervals and their corresponding linear function. For any $x$ in one interval, the corresponding $y$ calculated by its corresponding function is always less than $y$s calculated by other functions. For example, in Fig. 7, for any $x$ in $(-\infty, x_1]$, the line A will give minimum $y$, and for any $x$ in $[x_1,x_2]$, the $y$ obtained by line B is minimal. So if the intervals and their corresponding linear functions are calculated, binary search method can be used to determine which interval $x$ belongs to and the corresponding linear function will be used to calculate $y$. The computation cost of binary search is $O(logM)$, and $Q$ values will be updated, so the total computation cost is $O(Q×logM)$.

\begin{figure}
\begin{center}
\includegraphics[height=1.6in, width=1.6in]{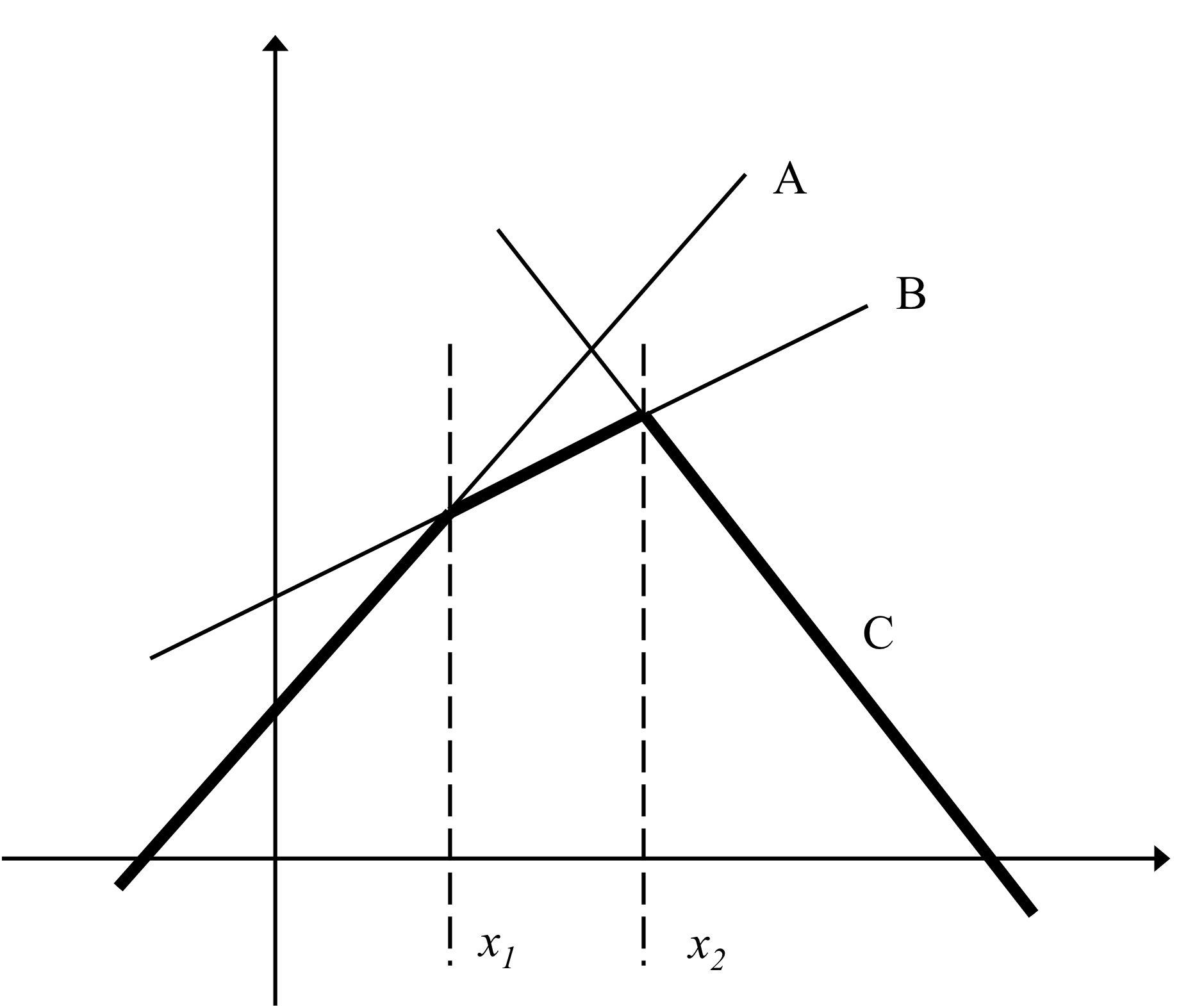}
\caption{Example of Intervals}
\end{center}
\end{figure}

To discuss the algorithm of calculating intervals and theirs corresponding linear functions, a data structure named ValidLinearFunction will be introduced. In the algorithm, ValidLinearFunction = $(f(x), x_l, x_u)$, where $f(x)$ denotes the linear function, $x_l$ denotes the lower bound of the interval, and $x_u$ denotes the upper bound. The interval of lines will be maintained when calculating. After a new line is added, we calculate it's interval and update previous lines' intervals by their intersection. As each line will be pushed or popped as most once, the time complexity is $O(M)$, and the computation cost of sorting slops of lines is $O(MlogM)$. So the total computation cost is $O(QlogM + MlogM + M)$.

In the beginning of dynamic programming algorithm, the slopes of lines $(-F(b_{k-1}))$ and the $x$'s ($cost(b_k)$) can be sorted once and used. So the time complexity can be reduced to  $O(Q+M)$.

\newpage
\bibliography{bin-design}
\bibliographystyle{aaai}
\end{document}